\documentclass[final]{elsart5p}

\usepackage{graphicx,amssymb}
\journal{Acta Materialia}

\begin{document}
\def\cna{\mbox{Co$_{38}$Ni$_{33}$Al$_{29}$}}

\begin{frontmatter}
\title{Microstructure of precipitates and magnetic domain structure in an annealed \cna\  shape memory alloy}
\author[Antwerp]{B. Bartova\corauthref{cor}\thanksref{now}},
\corauth[cor]{Corresponding author.}
\thanks[now]{Now at Swiss Federal Institute of Technology (EPFL), MXC 133 (Batiment MXC), Station 12, CH-1015  Lausanne, Switzerland, Tel.: +41 216934884, Fax: +41 216934401}
\ead{barbora.bartova@epfl.ch}
\author[Glasgow]{N. Wiese},
\ead{n.wiese@physics.gla.ac.uk}
\author[Antwerp]{D. Schryvers},
\author[Glasgow]{J.N. Chapman},
\author[Prague]{S. Ignacova}

\address[Antwerp]{EMAT, University of Antwerp, Groenenborgerlaan 171, B-2020 Antwerp, Belgium} 
\address[Glasgow]{Department of Physics and Astronomy, University of Glasgow, Glasgow G12 8QQ, United Kingdom}
\address[Prague]{Department of Metals, Institute of Physics, ASCR, v.v.i., Na Slovance 2, CZ-182 21 Prague 8, Czech Republic}

\begin{abstract}
The microstructure of a \cna\ FSMA was determined by conventional
transmission electron microscopy, electron diffraction studies
together with advanced microscopy techniques and in-situ Lorentz
microscopy. 10 to 60~nm sized rod-like precipitates of hcp
$\epsilon$-Co were confirmed to be present by HRTEM. The orientation
relationship between the precipitates and B2 matrix is described by
the Burgers orientation relationship. The crystal structure of the
martensite obtained after cooling is tetragonal L1$_{0}$ with a (1-11)
twinning plane. The magnetic domain structure was determined during an
in-situ cooling experiment using the Fresnel mode of Lorentz
microscopy. While transformation proceeds from B2 austenite to
L1$_{0}$ martensite, new domains are nucleated leading to a decrease
in domain width, with the magnetization lying predominantly along a
single direction. It was possible to completely describe the
relationship between magnetic domains and crystallographic directions
in the austenite phase though complications existed for the martensite
phase.
\end{abstract}


\end{frontmatter}

\section{Introduction}
\label{Intro}

\begin{figure} [ht]
\begin{center} 
\includegraphics{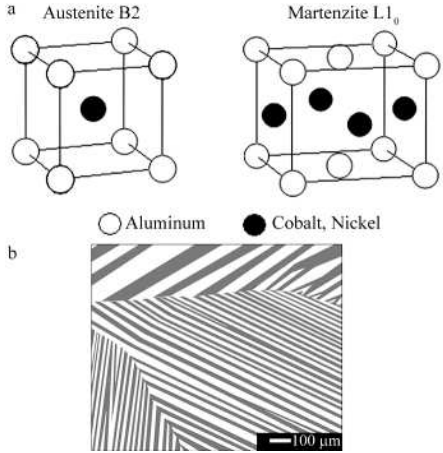}
\end{center} 
\caption{Schematic drawing of (a) the unit cells of the B2 austenite
  and the L1$_{0}$ martensite. In (b) a typical morphology of the
  microtwinned martensite plates is depicted based on an exemplary
  observation by Karaca et al.~\cite{Karaca2003}.}
\label{fig1} 
\end{figure}

Ferromagnetic shape memory alloys (FSMAs) are being intensively
studied because of their potential applications as smart materials.
Martensitic transformations and lattice reorientation processes in
FSMAs can be triggered not only by changes in temperature and stress,
as in conventional SMAs, but also by applying an external magnetic
field. To date, many such systems have been investigated, including
Ni$_{2}$MnGa \cite{Ullakko1996,Mogylnyy2003,LiuGD2004}, Ni$_{2}$MnAl
\cite{Fujita2000}, Fe$_{70}$Pd$_{30}$ \cite{James1998} and Fe$_{3}$Pt
\cite{Kakeshita2000}, all experiencing large strains induced by an
external magnetic field. Recently, the Co-Ni-Al system has received
increased interest as a new FSMA \cite{Tian1998,Oikawa2001,Oikawa2006}
since these alloys have low density, high melting point, good
corrosion resistance and high strength, even at temperatures as high
as 573~K. Moreover, the constituent elements are cheaper compared to
some other FSMAs (Fe-Pt, Fe-Pd). The Co-Ni-Al system undergoes a
martensitic transformation from $\beta$-phase (B2, cubic) austenite to
L1$_{0}$ (tetragonal) martensite in a temperature range between 93 and
393~K depending on composition with the symmetry loss being
responsible for the formation of microtwinned variants in the product
phase as shown in Fig.~\ref{fig1} \cite{Karaca2003}. It is this
martensitic transformation which is responsible for the shape memory
effect and pseudoelasticity. The single $\beta$-phase in
polycrystalline material is extremely hard and brittle, but the
presence of a secondary $\gamma$-phase, which has an Al disordered
face centered cubic structure \cite{Oikawa2001,Kainuma1996},
significantly improves the ductility \cite{Brown2005,Hamilton2005}.
Furthermore, the martensitic start temperature (TM$_{s}$) and Curie
temperature (T$_{c}$) can be independently controlled by the
composition. TM$_{s}$ decreases with increasing content of Co and Al
whereas T$_{c}$ increases with increasing Co content and decreasing
amounts of Al \cite{Oikawa2001}. Thus choice of the right composition,
annealing conditions and desired transition temperatures is necessary
to obtain a promising material for a wide range of applications. 

\par The purpose of this work is a detailed study of the
microstructure of austenite and martensite, going beyond existing work
\cite{Murakami2002}, and an investigation of the relation between
magnetic and crystallographic structure. Since TM$_{s}$ for the
material studied here is below room temperature, in-situ cooling
experiments were performed involving conventional transmission
electron microscopy (CTEM) and Lorentz microscopy \cite{Chapman1984}.

\section{Experimental procedures}
\label{Experiment}

A \cna\ alloy was obtained from Special Metals Corporation, New
Hartford, NY. The alloy was melted and single crystals were grown by
the directional Bridgman technique with a pulling rate of 10~$\mu$m.s$^{-1}$ 
in an alumina crucible at 1803~K in a vacuum. Annealing was
carried out at 1548~K for 4~hours in an Ar atmosphere followed by
quenching into ice water. The transformation temperatures are
determined to be 231~K for TM$_{s}$ and 266~K for the austenite start
temperature (TA$_{s}$) as measured by differential scanning
calorimetry (DSC). 

\par The phase composition was determined using a JEOL JSM 5510
scanning electron microscope (SEM) equipped with an INCA energy
dispersive X-ray (EDX) microanalysis system. TEM specimens were
prepared by twinjet electro-polishing in a 20\% sulfuric acid and 80\%
methanol electrolyte at 278~K \cite{Karaca2004}. Conventional TEM was
performed on a Philips CM20 and high resolution transmission electron
microscopy (HRTEM) images were acquired on a JEOL JEM 4000EX
microscope. The spectrometer used for energy filtered TEM (EFTEM)
measurements is a post-column GATAN Imaging Filter (GIF200) mounted
onto a 300keV Philips CM30 field emission gun (FEG) microscope. EFTEM
maps were obtained with the commercial software package Digital
Micrograph. Diffraction pattern simulations were carried out with the
commercial software package CrystalKitX.
\par To study the magnetic domain structure of the sample, the Fresnel
mode of Lorentz microscopy was used \cite{Chapman1984}. The TEM was a
modified Philips CM20 equipped with (non-immersion) Lorentz lenses,
thereby allowing magnetic imaging in a field-free environment with the
standard objective lens switched off \cite{Chapman1994}. All
experiments involving Lorentz microscopy were performed with an
untilted specimen.

\section{Results and discussion}
\label{ResDis}
\subsection{Co-rich precipitates in the austenite matrix}
\label{Precipitates}

\begin{figure} [ht]
\begin{center} 
\includegraphics{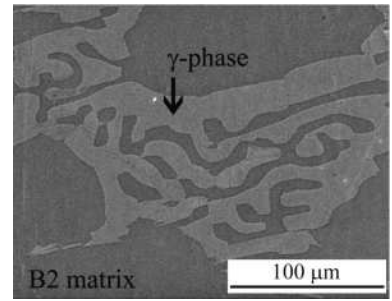}
\end{center} 
\caption{SEM image of the microstructure of \cna\ annealed alloy consisting of B2 matrix with dispersed $\gamma$-phase.} 
\label{fig2} 
\end{figure}

\begin{figure} [ht]
\begin{center} 
\includegraphics{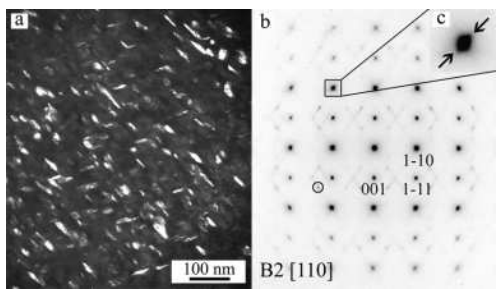}
\end{center} 
\caption{(a) Dark-field image of the Co precipitates present in the B2 matrix taken from the reflection marked with a circle. (b) Diffraction pattern taken from the [110] zone axis of B2 matrix (a=0.287~nm \cite{Karaca2004}). (c) Streaks around -11-3 reflection of the matrix marked by arrows.} 
\label{fig3} 
\end{figure}

The morphology of the sample, following annealing and subsequent
quenching, consists of the B2 matrix and a $\gamma$-phase. The
experimentally obtained phase compositions are given in
Table~\ref{tab1} with the microstructure shown in the SEM image,
Fig.~\ref{fig2}. Table~\ref{tab1} lists the elemental concentrations
measured by EDX, averaging over 7 measurements for the $\gamma$-phase
and 5 measurements for the B2 matrix, yielding the indicated standard
deviations. A dark field TEM image reveals small precipitates present
in the B2 matrix in Fig.~\ref{fig3}a. The weak reflection used to
obtain the dark field image is encircled in the corresponding selected
area diffraction pattern (SAED), Fig.~\ref{fig3}b. Streaks can be
detected around the spots of the matrix in SAED pattern shown in
Fig.~\ref{fig3}c. These streaks can arise because of modifications to
the shape of reciprocal lattice points originating from either the
shape of the crystal defects or lattice strain connected with them
\cite{Edington}. Both causes exist in the present case, but are
difficult to separate because they occur in the same direction.

\begin{table} [ht]
\begin{center}
\begin{tabular}{|cc|cc|}
\hline
At.[\%] & Al & Co & Ni \\\hline
$\gamma$-phase 	& 	$17.2\pm0.5$  &  $53.2\pm0.7$ & $29.6\pm0.2$ \\
B2 matrix     	&  	$29.2\pm0.2$  &  $38.1\pm0.2$ & $32.4\pm0.2$ \\\hline
\end{tabular}
\end{center}
\caption{Chemical composition of \cna\ alloy annealed at 1548~K/4~h
  measured by SEM EDX.}
\label{tab1} 
\end{table}

\par
\begin{figure} [ht]
\begin{center} 
\includegraphics{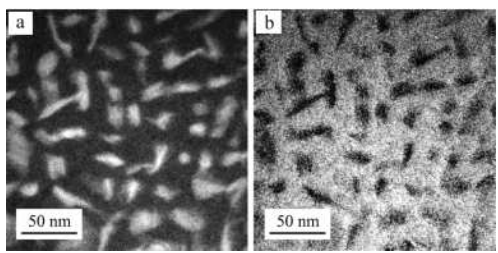}
\end{center} 
\caption{(a) EFTEM Co map and (b) Ni map revealing the Co enrichment in the precipitates.}
\label{fig4} 
\end{figure}

 The rod-like precipitates have dimensions ranging from 10 to 60~nm 
for the longest axis. EFTEM maps presented in Fig.~\ref{fig4}
indicate that these precipitates are enriched in Co with respect to
the matrix. Quantification of the Co content, however, was not
possible due to an inevitable and unknown overlap with the matrix.
Cobalt can exist in two crystalline forms, face-centered cubic
($\alpha$-Co; a=0.35441~nm) and close-packed hexagonal ($\epsilon$-Co;
a=0.25074~nm, c=0.40699~nm) with an allotropic transformation
occurring at 695~K \cite{Nash}.

\par
\begin{figure} [ht]
\begin{center} 
\includegraphics{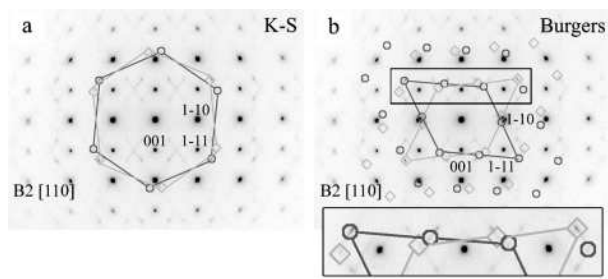}
\end{center} 
\caption{SAED pattern taken from the [110] zone axis orientation of B2 matrix with added simulation of orientation relationship for (a) $\alpha$-Co and (b) $\epsilon$-Co precipitates. Simulations of the 2 variants for fcc and hcp Co are given by open circles and squares connected by lines.} 
\label{fig5} 
\end{figure}

 Assuming each possibility in turn, the orientation relationships
between precipitates and matrix were simulated for both cubic and
hexagonal cases. Figure~\ref{fig5}a shows the reciprocal orientation
relationship between the body-centered cubic (bcc) lattice of the B2
matrix and 2 variants of fcc Co precipitates, the former being
observed in the [110] zone orientation:
\begin{center}
	\begin{tabular*}{1.0\textwidth}{ll}
 	 	Variant 1: & (110)$_{B2}$//(111)$_{fcc}$, [-11-1])$_{B2}$//[-110]$_{fcc}$\\
  	Variant 2: & (110)$_{B2}$//(111)$_{fcc}$, [-111])$_{B2}$//[-110]$_{fcc}$
	\end{tabular*}
\end{center}
These match the Kurdjumow-Sachs (K-S) orientation relationship typical
for fcc-bcc solid systems. The potential Co reflections are given by 
open circles and squares, respectively, connected by lines, present to indicate clearly both
variants. From this schematic it is clear that several Co-reflections
cannot be explained by the present simulation, although some
unexplained ones could arise from double diffraction.
Figure~\ref{fig5}b illustrates a comparable simulation for the
orientation relationship between the bcc lattice of the B2 matrix and
2 variants of hcp Co precipitates:
\begin{center}
	\begin{tabular*}{1.0\textwidth}{ll}
  	Variant 1: & (110)$_{B2}$//(001)$_{hcp}$, [-11-1])$_{B2}$//[110]$_{hcp}$\\ 
  	Variant 2: & (110)$_{B2}$//(001)$_{hcp}$, [-111])$_{B2}$//[110]$_{hcp}$
	\end{tabular*}
\end{center}
These correspond to the Burgers orientation relationship for hcp-bcc
solid systems. Here there is better agreement between simulated and
experimental SAED patterns, as can be seen from the enlargement in figure~\ref{fig5}b 
where small but distinct reflections show up on all simulated
positions.

\begin{figure} [ht]
\begin{center} 
\includegraphics{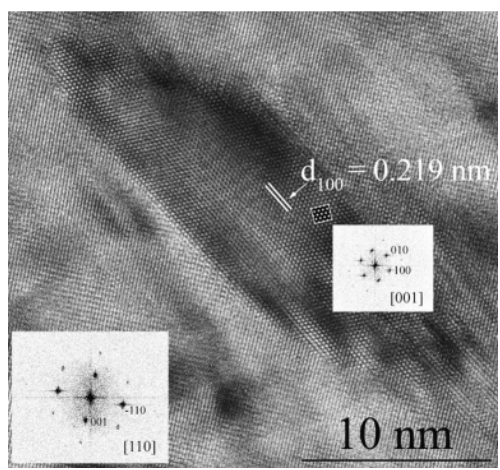}
\end{center} 
\caption{HRTEM image shows hcp precipitate in B2 matrix plus the FFT plots indicate the bcc matrix in a [110] and hcp precipitate in a [001] zone axis orientation. The simulated image of hcp [001] is added as an outlined input.} 
\label{fig6} 
\end{figure} 

\par
\begin{figure} [ht]
\begin{center} 
\includegraphics{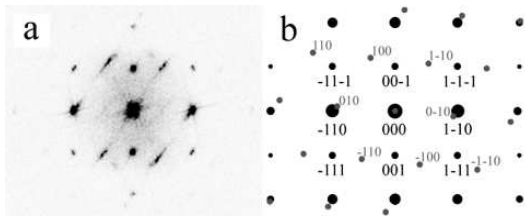}
\end{center} 
\caption{(color online) (a) FFT plot taken from whole area of Fig.~\ref{fig6}. (b) Simulated diffraction pattern combining $\epsilon$-Co precipitate (variant 1) along the [001] zone axis with B2 matrix in the [110] zone.} 
\label{fig7} 
\end{figure}

 In the high resolution image of Fig.~\ref{fig6}, a single
precipitate in the B2 matrix can be seen. From the fast Fourier
transform (FFT) pattern it can be concluded that the B2 matrix is
viewed along a [110] zone axis. The FFT of the precipitate clearly
reveals hexagonal [001] or cubic $<$111$>$ symmetry. From the
tabulated d-spacings given in Table~\ref{tab2} it can be concluded
that the measured value d$_{100}$=0.219~nm fits with the values for
hcp $\epsilon$-Co. Figure~\ref{fig7}a shows the FFT pattern from the
entire area of Fig.~\ref{fig6}. It corresponds to variant 1 of the
Burgers orientation relationship which is in good agreement with
simulation for the same case shown in Fig.~\ref{fig7}b. This result is
in accordance with conclusions reached by Tian et al. \cite{Tian1998},
although the alloy composition and annealing conditions were
different. The stability and transformation temperature of the two
allotropic modifications of Co depend on the grain size, purity and
degree of lattice distortion both before and after heat treatment
\cite{Taylor1950}. Owen and Jones concluded that in the very small
grains the stable structure is fcc up to a temperature of at least 873~K 
whereas in larger grains hcp structure is stable. However, specimens
with different grain sizes annealed up to 1273~K and subsequently
quenched into water showed a mixture of both structures
\cite{Owen1954}. In the present study the sample was homogenized at
1548~K for 4 hours followed by quenching into iced water. Such rapid
cooling leads to non-equilibrium solidification and can explain the
presence of precipitates with hexagonal structure, even though the
phase diagram predicts that the cubic structure is stable at
temperatures higher than 693~K. In the investigated material only
precipitates with hexagonal structure were found although the as cast
material contains fcc $\alpha$-Co particles with nanometer size \cite{Bartova2007}.

\begin{table} [ht]
\begin{center}
\begin{tabular}{cc|cc}
$\alpha$-Co & fcc & $\epsilon$-Co & hcp \\\hline
$\mbox{d}_{hkl}$ (nm)  &  (hkl)  & $\mbox{d}_{hkl}$ (nm) & (hkl) \\\hline
				0.2506 & (-110) (011) (-101) & 0.2171 & (-110) (010) (100) \\
          		& (110) (0-11) (101)  & 0.2035 & (002) \\             
\end{tabular}
\end{center}
\caption{d-spacings for cubic $\alpha$-Co and hexagonal
  $\epsilon$-Co precipitates; with the latter corresponding to the
  measured $\mbox{d}_{100}=0.219$~nm value.}
\label{tab2} 
\end{table}

\par The present observations of $\gamma$-phase and
Co-rich nanoscale precipitates indicate that the matrix will contain a
lower amount of Co than the nominal composition of the material, which
will have its influence on the structural and magnetic transformation
parameters of the bulk matrix.

\subsection{Relationship between magnetic and crystallographic structure}
\label{Magneticdomains}
\begin{figure} [ht]
\begin{center} 
\includegraphics{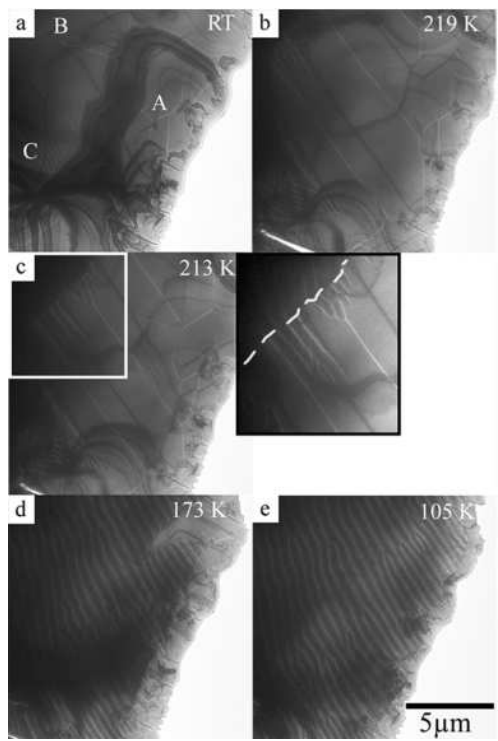}
\end{center} 
\caption{Fresnel images of the magnetic domain structure observed during an in-situ cooling experiment. (a-b) The material keeps the high temperature structure (austenite). (c) The low temperature phase (martensite) starts to develop in the left upper corner; the habit plane between austenite and martensite can be seen in the enlargment. (d-e) The material is fully transformed to martensite.} 
\label{fig8} 
\end{figure}

Figure~\ref{fig8}a shows a Fresnel image of the sample obtained at
room temperature. The existence of magnetic domain wall contrast
indicates that the sample is ferromagnetic at this temperature. A
subsequent heating experiment to 363~K, the maximum temperature
accessible with the equipment used, revealed negligible change in the
domain wall contrast. Therefore, it can be assumed that the Curie
temperature is well above room temperature, which is substantially
higher than the value measured by Murakami et al. in the same alloy
homogenized at 1623~K \cite{Murakami2002}, a difference possibly due
to a difference in secondary phase formation and thus in net matrix
composition.

\par In the thinner part of the sample close to the edge
(region A), large domains with an irregular domain wall pattern are
observed, indicating a low crystallographic anisotropy within the
austenite phase. In thicker areas, one also finds two areas with
regularly spaced domains, labeled B and C in Fig.~\ref{fig8}a. Each of
these areas show a uniform periodicity and direction, and they are
probably related to residual traces of the martensitic phase from
preceding cooling experiments.
\par A set of Fresnel images taken at different temperatures during an
in-situ cooling experiment is presented in Fig.~\ref{fig8}b-e. Below
approximately 223~K the magnetic contrast arising from the areas with
small period domains as well as the extent of them decreases, and
ordering of the magnetic domain structure into micrometer sized
domains with an average width of approximately 1.9~$\mu$m is observed
over much of the field of view and the crystal structure remains that
of austenite. In Fig.~\ref{fig8}c at 213~K the martensite habit plane appears in the upper left corner as indicated by the dashed line. At the same time the magnetic domain structure starts to change and approximately parallel domain walls with a significantly reduced domain width of about 0.3~$\mu$m appear. This change in domain structure is related to the structural transformation from austenite to martensite as discussed below in detail. The TM$_{s}$ observed during the in-situ cooling experiment is lower than
the TM$_{s}$ measured by DSC, the explanation being the retarding effect of the thin foil with respect to the bulk material.
Upon further cooling the small period magnetic domain
structure spreads into the whole field of view, consistent with a
complete transformation to the martensite phase (Fig.~\ref{fig8}d,e).
Note that as well as there being some variation in period, the
orientation of the domain walls varies by ${\sim}$45$^{\circ}$ across
the image.

\par
\begin{figure} [t]
\begin{center} 
\includegraphics{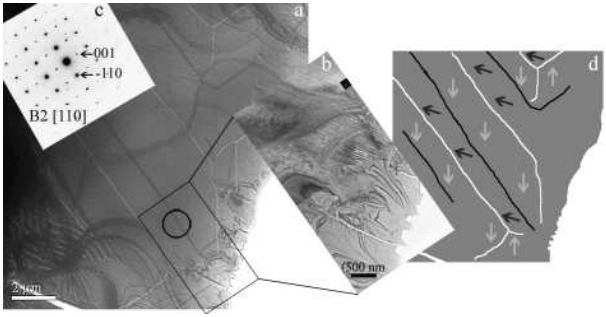}
\end{center} 
\caption{(a) Magnetic domain structure at 219~K observed by Lorentz microscopy; (b) The same area of the specimen observed by CTEM at 233~K; (c) Selected area diffraction pattern (SAED) taken from the region marked by the circle. (d) Model of the components of magnetization projected on the sample plane.}
\label{fig9} 
\end{figure}

 In order to relate the orientation of the magnetic domains with
the crystallographic directions of the underlying structures, the
in-situ cooling experiment was repeated in a conventional transmission
electron microscope. The same area of the sample was imaged in both
cases, identified by recognizable features at the edge of the sample.
Since the magnetization distribution is irregular at room temperature,
the image of magnetic domains at 219~K is compared with bright field
images taken at 233~K. Figures ~\ref{fig9}a and b present the
comparison of Lorentz and CTEM microscopy. Whilst, as noted above, the
specimen was untilted during the Lorentz microscopy experiments, a
small tilt of $\approx$10$^{\circ}$ was introduced to allow the bright
field image to be recorded with the electron beam parallel to the
nearest low index zone axis of the B2 matrix. The inset,
Fig.~\ref{fig9}c, shows the SAED originating from the encircled area,
and the zone axis indexes as [110]. Comparison of the diffraction
pattern and the magnetic image shows that the majority of walls run
parallel to [-110], although walls parallel to [001] and [-11-1] can
also be observed. This in turn suggests that domain walls of differing
angle are present as shown in the schematic, Fig.~\ref{fig9}d.
Generically similar structures have been seen in Fe whiskers and in
thin epitaxial Fe films \cite{Gu1995,Daboo1995}. From the schematic it
can be seen that the easy axes of magnetization are $<$111$>$. Thus an
internally consistent picture has been constructed. Given that certain
of the $<$111$>$ directions lie close to but not exactly in the plane
of the thinned TEM sample, evidenced by the need to tilt the specimen
through a small angle to make the electron beam coincide with the zone
axis, we would expect the specimen to form a domain structure rather
than exist in a single domain state. However, given that the easy axes
do not lie much out of plane, only modest magnetic surface charge
arises there and the resulting domain structure has a comparatively
large periodicity, this being determined by a balance between domain
wall and magnetostatic energy \cite{Hubert}. Moreover, the fact that
more domain walls run parallel to [-110] than to [001] is simply a
consequence of the (uncontrolled) displacement of the foil normal from
the [110] direction.

\begin{figure} [ht]
\begin{center} 
\includegraphics{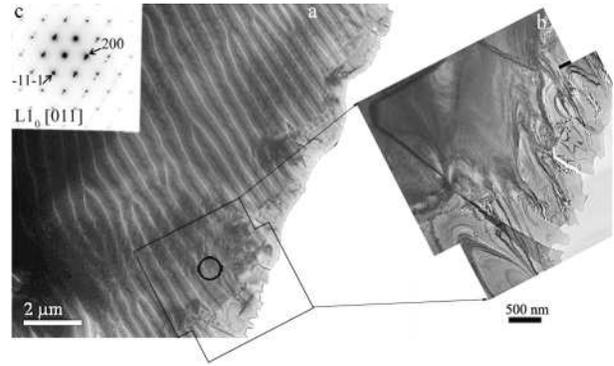}
\end{center} 
\caption{(a) Magnetic domain structure at 105~K observed by Lorentz microscopy; (b) The same area of the specimen observed by CTEM at 113~K; (c) Selected area diffraction pattern (SAED) taken from the region marked by the circle.}
\label{fig10} 
\end{figure} 

\par The same procedure for analysis was repeated after cooling the
specimen to 113~K to determine more fully the magnetic structure in
the L1$_{0}$ martensite phase. In this case, the specimen had to be
tilted through an angle $\approx$25$^{\circ}$ to make the electron
beam parallel to a prominent zone axis. Figure~\ref{fig10} shows
relevant images together with the SAED, the latter indexed as [011].
Around the small area from which the SAED was recorded the orientation
of the domain walls is close to [11-1] but, as already noted, there is
a considerable spread of wall orientation across the image, roughly
speaking it varies from [11-1] to [02-2]. Some care is needed here in
that we do not have diffraction information for other parts of the
imaged region, but there is no evidence of any grain boundaries being
present. However, local twinning does exist and evidence for it can be
seen in the diffraction pattern where only the row of diffraction
spots running through [-11-1] and [1-11] is unsplit. Hence the
situation is considerably more complex than for the
austenite.
\par Despite this, in many senses the domain structure is
simpler than in austenite in that the martensite appears to have
uniaxial anisotropy, albeit with an axis that varies about some mean
direction, and that is inclined at a significant angle to the foil
surface. We make these assertions with confidence by comparison with
earlier work where very similar structures were observed in thinned
sections through differently oriented Nd-Fe-B grains
\cite{Young1993,Yi2000}. However, identification of a preferred
crystallographic axis is inappropriate here in that several factors,
none unique to the material but to an extent dependent on the specific
area of sample investigated, play important roles. These are the
orientation of the surface normal of the foil and the relative
contributions made by the untwinned and twinned parts of the specimen.
It is reported in the literature \cite{Fujita2003} that the martensite
phase has uniaxial anisotropy, [001] being the hard axis. The [001]
axis here is inclined at 50.8$^{\circ}$ to the zone axis and were that the only
consideration presumably the magnetisation would lie in the plane of
the foil along the line of intersection of the foil surface and the
plane perpendicular to [001]. The complication arises due to the
extensive twinning within the crystal and the fact that angle between
the c-axis of the twinned and untwinned parts is 66$^{\circ}$. Hence there is
no common direction in the plane of the specimen and the preferred
magnetisation orientation, determined by the minimisation of the sum
of the anisotropy and magnetostatic energies, is for this area of
sample, inclined to the specimen surface. This in turn means that a
uniaxial domain structure with 180$^{\circ}$ domain walls running parallel to
the projection of the net easy axis into the plane of the specimen is
favoured.

\begin{figure} [ht]
\begin{center} 
\includegraphics{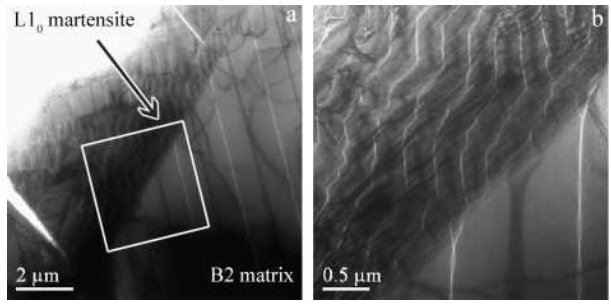}
\end{center} 
\caption{(a) Fresnel image of a partially transformed region of the specimen. (b) Detail close to the interface between austenite and martensite as indicated.}
\label{fig11} 
\end{figure} 

\par In some regions close to the hole, the phase
transformation is incomplete. At these positions, the habit plane
between the austenite and martensite regions can be observed. In
Fig.~\ref{fig11}, a Fresnel image of such a region is shown. The
magnetic domain structure has to accommodate across the interface, an
inevitable consequence of the difference between the preferred domain
spacings in the two phases. Moreover, in this very thin region, it is
clear that there is heavy twinning within the martensite and locally
the domain walls adopt irregular zig-zag structures with walls running
for variable distances along either of two preferred directions. This
behaviour is not observed in the thicker part of the specimen but is
consistent with the earlier observation that there are a number of
factors contributing to the orientation of domains in martensite.
Furthermore, it has been observed in a Fe$_{68.8}$Pd$_{31.2}$ FSMA with very thin
martensite plates, that the magnetostatic energy plays a major role
determining the magnetization directions \cite{Murakami2006}.
Returning to the interface itself, it appears that the domain
structure in the austenite, where the domain walls are more widely
separated is the more affected. Specifically, the wall structure
appears to change within 100-200~nm of the interface with small
triangular regions developing. The most likely explanation for this is
to rotate the local magnetisation to run more closely parallel to the
interface, thereby lowering the probability of magnetisation vectors
from the two phases meeting head on, a situation in which the
magnetostatic energy is maximised. Similar occurrences were observed
at the boundaries between differently oriented grains in thinned
sections of sintered Nd-Fe-B \cite{Young1993}.

\section{Conclusions}
\label{Concl}
The microstructure of an annealed (1548~K/4~h) \cna\ alloy was
investigated by transmission electron microscopy. In addition to the
major constituents (B2 matrix and $\gamma$-phase), rod-like
precipitates ranging from 10 to 60~nm of $\epsilon$-Co were observed
in the austenite phase. EFTEM measurements confirmed that the
precipitates were Co-rich. The orientation relationship between the
precipitates and the B2 matrix was found to be the Burgers orientation
relation. The martensite crystal structure is tetragonal L1$_{0}$ with
a (1-11) twinning plane. Quite different magnetisation distributions
were observed in the two phases, reflecting the different crystal
structures. Whilst the domain structures are influenced by the fact
that they were observed in a thin section of material where shape
anisotropy plays a major role, it was possible to reconcile the domain
configurations with the known crystallography. A comparatively simple
relation was found in austenite whereas the observations in martensite
were complicated by the presence of the extensive twinning in the
sample.

\section{Acknowledgments}
\label{Aknowl}
B. Bartova, N. Wiese and S. Ignacova gratefully acknowledge financial support from the MULTIMAT Marie Curie Research Training network (MRTN-CT-2004-505226).

\end{document}